\documentclass[11pt]{article}
\usepackage{amsmath,amsfonts,amssymb}
\usepackage{fullpage}
\usepackage{setspace}
\usepackage{graphicx}
\usepackage{color}
\usepackage{bm}
\usepackage{upgreek}
\usepackage{comment}
\usepackage{authblk}

\begin{document}

\title{\bf Unraveling the behavior of the individual ionic activity coefficients on the basis of the balance of ion-ion and ion-water interactions}

\author[1]{M\'onika Valisk\'o}
\author[1]{Dezs\H{o} Boda}
\affil[1]{Department of Physical Chemistry, University of Pannonia,  P. O. Box 158, H-8201 Veszpr\'em, Hungary}

\date{\today}

\maketitle

\begin{abstract}
We investigate the individual activity coefficients of pure 1:1 and 2:1 electrolytes using our theory that is based on the competition of ion-ion (II) and ion-water (IW) interactions (Vincze {\it et al.}, J. Chem. Phys. 133, 154507, 2010).
The II term is computed from Grand Canonical Monte Carlo simulations on the basis of the implicit solvent model of electrolytes using hard sphere ions with Pauling radii.
The IW term is computed on the basis of Born's treatment of solvation using experimental hydration free energies.
The two terms are coupled through the concentration-dependent dielectric constant of the electrolyte.
With this approach we are able to reproduce the nonmonotonic concentration dependence of the mean activity coefficient of pure electrolytes qualitatively without using adjustable parameters. 
In this paper, we show that the theory can provide valuable insight into the behavior of individual activity coefficients too.
We compare our theoretical predictions against experimental data measured by electrochemical cells containing ion-specific electrodes.
As in the case of the mean activity coefficients, we find good agreement for 2:1 electrolytes, while the accuracy of our model is worse for 1:1 systems.
This deviation in accuracy is explained by the fact that the two competing terms (II and IW) are much larger in the 2:1 case so errors in the two separate terms have less effects.
The difference of the excess chemical potentials of cations and anions (the ratio of activity coefficients) is determined by asymmetries in the properties of the two ions: charge, radius, and hydration free energies.
\end{abstract}

\begin{center}
\end{center}

\newpage
\begin{center}
\textbf{TOC graphic:}\\
\vspace{0.5cm}
\rotatebox{0}{\scalebox{0.8}{\includegraphics*{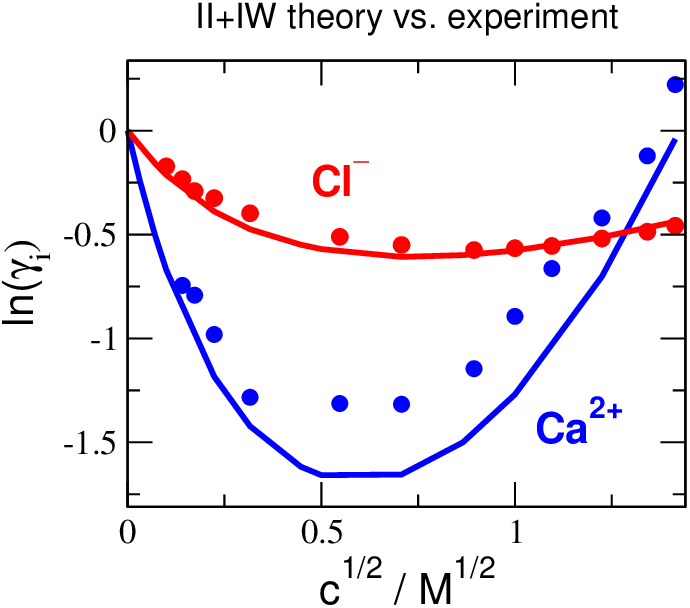}}}
\end{center}


\newpage


\section{Introduction}
\label{sec:intro}

The individual activity coefficient, $\gamma_{i}$, of an ionic species $i$ in an electrolyte solution describes the deviation from ideality through the excess chemical potential
\begin{equation}
\mu_{i}^{\mathrm{EX}} =  kT \ln \gamma_{i}
\end{equation}
that is defined by
\begin{equation}	
\mu_{i}=\mu_{i}^{0}+kT\ln c_{i}+\mu_{i}^{\mathrm{EX}} ,
\end{equation} 
where $\mu_{i}$ is the chemical potential of species $i$, $c_{i}$ is its concentration, $\mu_{i}^{0}$ is a reference chemical potential independent of the concentration, $\mu_{i}^{\mathrm{EX}}$ is the excess chemical potential characterizing the effect of interaction between particles, $k$ is Boltzmann's constant, and $T$ is the temperature.
The reference point is chosen in such a way that $\lim_{c\rightarrow 0}\mu_{i}^{\mathrm{EX}} = 0$, where $c$ is the salt concentration \cite{lewis1923}.
The salt concentration is defined as $c=c_{+}/\nu_{+}=c_{-}/\nu_{-}$ with $\nu_{+}$ and $\nu_{-}$ being the stoichiometric coefficients of the cation and the anion in a simple electrolyte with the stoichiometry 
\begin{equation}
\mathrm{C}_{\nu_{+}} \mathrm{A}_{\nu_{-}} \rightleftarrows \nu_{+} \mathrm{C}^{z_{+}} + \nu_{-} \mathrm{A}^{z_{-}},
\end{equation} 
where C and A refer to cations and anions, while $z_{+}$ and $z_{-}$ are the valences of the ions.

The individual activity coefficient is an important quantity for several reasons.
This quantity appears in the Nernst-equation for the electrode potential, $E$, in a half-cell, for example, 
\begin{equation}
 E = E^{0} + \dfrac{kT}{z_{i}e}\ln (\gamma_{i} c_{i}) ,
\label{eq:nernst}
\end{equation} 
where $e$ is the charge of the proton and $E^{0}$ is a standard electrode potential.
Knowledge of the individual activity coefficient, furthermore, is crucial in many biological and technological phenomena/processes such as Donnan equilibrium, ion transport, ion exchange, and corrosion.

The individual activity coefficient, however, is not readily available from measurements.
We usually do not know its value, therefore, it is common to use the mean activity coefficient in Eq.\ \ref{eq:nernst} instead.
The mean activity coefficient is defined as
\begin{equation}
 \gamma_{\pm} = \gamma_{+}^{\nu_{+}/\nu}\gamma_{-}^{\nu_{-}/\nu} ,
\end{equation} 
where $\nu=\nu_{+}+\nu_{-}$.
Accordingly, the mean excess chemical potential is computed as
\begin{equation}
 \mu_{\pm}^{\mathrm{EX}} = \dfrac{\nu_{+}}{\nu} \mu_{+}^{\mathrm{EX}} + \dfrac{\nu_{-}}{\nu} \mu_{-}^{\mathrm{EX}}.
\end{equation} 
The mean quantities, $\gamma_{\pm}$ and $\mu_{\pm}^{\mathrm{EX}}$, can be measured accurately \cite{robinson-stokes,bockris-reddy,fawcett-book}.

The individual excess chemical potentials of the two ions, however, are not equal.
To what degree are they different is an important question that was addressed both in theoretical and experimental studies.
To characterize this, let us introduce the difference of the excess chemical potentials of the cation and the anion:
\begin{equation}
 \Delta \mu^{\mathrm{EX}} = \mu_{+}^{\mathrm{EX}}- \mu_{-}^{\mathrm{EX}} = kT \ln\left(  \dfrac{\gamma_{+}}{\gamma_{-}}\right) .
 \label{eq:delta_mu}
\end{equation} 
Note that once we have the mean and the difference, the individual excess chemical potentials can be calculated as
\begin{eqnarray}
 \mu^{\mathrm{EX}}_{+} = \mu^{\mathrm{EX}}_{\pm} + \dfrac{\nu_{-}}{\nu} \Delta\mu^{\mathrm{EX}} \nonumber \\
 \mu^{\mathrm{EX}}_{-} = \mu^{\mathrm{EX}}_{\pm} - \dfrac{\nu_{+}}{\nu} \Delta\mu^{\mathrm{EX}} .
 \end{eqnarray} 
Estimating the order of magnitude of $\Delta \mu^{\mathrm{EX}}$ is important for judging the error in Eq.\ \ref{eq:nernst} introduced by using the mean activity coefficient instead of the individual one.

An important experimental fact is that $\mu^{\mathrm{EX}}_{\pm}$ shows a non-monotonic concentration dependence: increasing the concentration from zero, it decreases from zero with a slope obeying the Debye-H\"uckel (DH) \cite{debye-huckel} limiting law, reaches a minimum at a large concentration, then increases again as the concentration approaches saturation. 
This phenomenon has been addressed by several studies starting from various empirical modifications \cite{robinson-stokes,bockris-reddy,fawcett-book,abbas-fpe-07,abbas-jpcb-09,fraenkel_mp_2010} of the DH theory \cite{debye-huckel}.
To first degree, these modifications tried to take the finite size of ions into account.

More developed statistical mechanical theories use a microscopic model, where the interactions acting between the particles of the system are represented with classical pair potentials.
In particular, most of the studies used the Primitive Model (PM) of electrolytes, where the ions are modeled as charged hard spheres (HS), while the solvent is modeled as a dielectric continuum with a dielectric constant $\epsilon(c)$ (in this work, we allow it to be concentration dependent).
The pair-potential describing this interaction can be given as
\begin{equation}
 u^{\mathrm{PM}}_{ij}(r)
=\left\{
        \begin{array}{ll}
    \infty & \; \mbox{for} \; \;  r<R_{i}+R_{j}\\
        \dfrac{z_{i}z_{j}e^{2}}{4\pi \epsilon_{0}\epsilon(c) r} & \; \mbox{for} \; \; r \geq R_{i}+R_{j}  ,
        \end{array}
        \right. 
\label{eq:pm}
\end{equation} 
where $R_{i}$ is the radius of ionic species $i$, $\epsilon_{0}$ is the permittivity of vacuum, and $r$ is the distance between the ions.
This model was applied in calculations using the mean spherical approximation (MSA) \cite{blum-mp-30-1529-1975,triolo-1-jpc-76,triolo-3-jcp-77,triolo-2-jpc-78,simonin-jpc-96,simonin-jpcb-97,fawcett-jpc-96,tikanen-bbgs-96,tikanen-jeac-97,lu-fpe-93,lopezperez_jeac_2000}, other theories \cite{outhwaite_jcsft_1991,molero_jcsft_1992,outhwaite_jcsft_1993,lu-fpe-93,inchekel}, and computer simulations\cite{valleau-jcp-80,abbas-fpe-07,abbas-jpcb-09}. 
This model is used in this paper to describe ion-ion interactions.

Two parameters of the PM (Eq.\ \ref{eq:pm}) have central importance in our discussion.
The ionic radius, $R_{i}$, is a molecular parameter, while the dielectric constant, $\epsilon(c)$, is a macroscopic (thermodynamic) quantity, that describes the screening of the environment of the ion.

In our previous works \cite{vincze-jcp-133-154507-2010,vincze-jcp-134-157102-2011,valisko-jcp-140-234508-2014}, we have introduced the II+IW model in which the excess chemical potential is split into two terms
\begin{equation}
 \mu_{i}^{\mathrm{EX}}=\mu_{i}^{\mathrm{II}}+\mu_{i}^{\mathrm{IW}} ,
\label{eq:ii+iw}
\end{equation}  
where II and IW refer to ion-ion and ion-water interactions, respectively.
The II term can be calculated on the basis of the PM using, for example, simulations (see Section \ref{sec:calc-II}).
The IW term was approximated through the Born-energy as described in Section \ref{sec:iw} in detail.
An important feature of our model is that it does not use adjustable parameters. 
Although it would be possible, we avoided this strategy because we were curious whether the nonmonotonic behavior of $\mu_{\pm}^{\mathrm{EX}}(c)$ can be understood and reproduced qualitatively without fitting. 
The parameters used in our model have strong experimental basis:
\begin{itemize}
\item The Pauling radii, $R_{i}$, are used in Eq.\ \ref{eq:pm} to compute the II term (see Table \ref{tab:ions}).
\item Experimental hydration free energies of the ions, $\Delta G_{i}^{\mathrm{s}}$, or, equivalently, the Born radii, $R_{i}^{\mathrm{B}}$, are used in the IW term (see Table \ref{tab:ions}). 
\item Experimental concentration-dependent dielectric constant, $\epsilon(c)$, is used in both the II and IW terms (see Table \ref{tab:salts}). This is a crucial quantity that establishes the coupling between the II and IW terms.
\end{itemize}
The way these parameters are handled indicates the difference between our approach and earlier studies.

Most of the earlier works fitted (increased) the radii of the ions (usually the radius of the cation) to obtain agreement with experiments \cite{triolo-1-jpc-76,triolo-3-jcp-77,triolo-2-jpc-78,simonin-jpc-96,simonin-jpcb-97,fawcett-jpc-96,tikanen-bbgs-96,tikanen-jeac-97,lopezperez_jeac_2000,inchekel,abbas-fpe-07,abbas-jpcb-09}. 
It was said that the increased ``solvated ionic radius'' took solvation into account by also including the hydration shell of tightly connected and oriented water molecules around the ion.
We criticized the idea of the ``solvated radius'' in our previous papers \cite{vincze-jcp-133-154507-2010,vincze-jcp-134-157102-2011,valisko-jcp-140-234508-2014} and pointed out that important configurations corresponding to cations and anions in contact are excluded from the statistical sample with this artificial concept.

It is an experimental fact that the dielectric constant of the electrolyte solution decreases with increasing concentration \cite{hasted_jcp_1958,giese_jpc_1970,barthel_1970,pottel,helgeson_ajs_1981,wei_jcp_1990,barthel_pac_1991,wei_jcp_1992,nortemann_jpca_1997,buchner_jpca_1999} due mainly to dielectric saturation \cite{anderson_cpl_1988,alper-jpc-90,smith_jcp_1994,gavish_arxiv_2012,levy_prl_2012,levy_jcp_2013,maribo_jpcb_2013a,renou_jpcb_2014,sega_jcp_2014}. 
The increasing electric field produced by the ions in more concentrated solutions orients the water molecules thus decreasing their ability to adjust their orientation in the solvation shell of an ion. 
Consequently, the screening ability of the solvent (expressed by its dielectric constant) decreases as the concentration of ions increases.
Some of the earlier works used a dielectric constant changing with concentration (either fitted \cite{triolo-3-jcp-77,simonin-jpc-96,lopezperez_jeac_2000} or experimental \cite{fawcett-jpc-96,tikanen-bbgs-96,tikanen-jeac-97}), but they ignored the change in the solvation free energy that should be included once an ion gets from an infinitely dilute solution (the reference state) to a concentrated solution (a different dielectric environment).
The IW term, therefore, were ignored by most of the authors. 
Two notable exceptions are the papers of Abbas \textit{et al.}\cite{abbas-fpe-07} and Inchekel \textit{et al.}\cite{inchekel} who took the IW interactions into account. 
These works were discussed in our previous papers \cite{vincze-jcp-133-154507-2010,valisko-jcp-140-234508-2014} in detail.

\begin{table}[t]
\renewcommand{\arraystretch}{1}
\begin{center}
\begin{tabular}{l|clcl}
\hline
Ion & \hspace{0.2cm}$z_{i}$\hspace{0.2cm} & $R_{i} /${\AA} & $\Delta G_{i}^{\mathrm{s}}$/kJmol$^{-1}$ & $R_{i}^{\mathrm{B}} /${\AA}  \nonumber \\
\hline
Li$^{+}$ & 1 & 0.6  & -529 & 1.3   \\
Na$^{+}$ & 1 & 0.95 & -424 & 1.62   \\
K$^{+}$ & 1 & 1.33  & -352 & 1.95   \\
Rb$^{+}$ & 1 & 1.48 & -329 & 2.084   \\
Cs$^{+}$ & 1 & 1.69 & -306 & 2.24   \\
\hline                              
Mg$^{2+}$ & 2 & 0.65 &-1931 & 1.42   \\
Ca$^{2+}$ & 2 & 0.99 &-1608 & 1.71   \\
Ba$^{2+}$ & 2 & 1.35 &-1352 & 2.03   \\
\hline                              
F$^{-}$ & -1 & 1.36  & -429 & 1.6  \\
Cl$^{-}$ & -1 & 1.81 &-304  & 2.26   \\
Br$^{-}$ & -1 & 1.95 &-278  & 2.47   \\
I$^{-}$ & -1 & 2.16  &-243  & 2.82   \\
\hline
\end{tabular}  
\caption{Experimental parameters of ions studied in this work: the valence, $z_{i} $, the Pauling radius \cite{pauling}, $R_{i}$, the hydration Hemholtz free energy \cite{fawcett-book}, $\Delta G_{i}^{\mathrm{s}}$, and the Born radius, $R_{i}^{\mathrm{B}}$ (computed from $\Delta G_{i}^{\mathrm{s}}$ on the basis of Eq.\ \ref{eq:bornw} with $\epsilon_{\mathrm{w}}=78.45$).}
\label{tab:ions}
\end{center}
\end{table}

Our earlier works \cite{vincze-jcp-133-154507-2010,vincze-jcp-134-157102-2011,valisko-jcp-140-234508-2014}, however, considered the mean excess chemical potential.
In this paper, we report results for the individual excess chemical potentials.
We compare our results to experimental data despite the fact that these experiments are not as well-established as those for the mean. 
A large portion of experimental data has been measured by the group of Vera and Wilczek-Vera (VWV) \cite{wilczekvera_fpe_1990,khoshkbarchi_aiche_1996,marcosarroyo1_jsc_996,rabie_jsc_1999,taghikhani_fpe_1999,taghikhani_cjce_2000,rodil_fpe_2001,rodil_aichej_2001,rodil_fpe_2003a,rodil_fpe_2003b,wilczekvera_aiche_2004,wilczekvera_fpe_2005,wilczekvera_fpe_2006b,wilczekvera_fpe_2006a,wilczekvera_ces_2007,wilczekvera_iecr_2009,wilczekvera_ces_2010,wilczekvera_fpe_2011,vera_jctd_442,wilczekvera_jctd_449}.
These measurements are strongly debated by other experimentalists \cite{malatesta_jsc_2000,malatesta_fpe_2005,malatesta_fpe_2006,malatesta_fpe_2010,malatest_ces_2010,zarubin_jctd_2011,zarubin_jctd_445,zarubin_jctd_451}.
Being theorists, we do not feel ourselves competent to judge in this debate; we just show the data.
Experiments from other sources are also available \cite{shatkay_ac_1969,hurlen_acssa_1979a,hurlen_acssa_1979b,hurlen_acssa_1981a,lee_jcice_2002,schneider_cit_2003,schneider_jsc_2004,zhuo_fpe_2008,sakaida_jpcb_2011,dash_isrn_2012}.
The experimental issues are briefly discussed in Section \ref{subsec:exp}.

\section{Previous works on the individual activity coefficients}
\label{sec:previouswork}

\subsection{Theory}
\label{sec:prevtheory}

Perhaps due to the confusion about the measurability and even the pure existence of the individual activity coefficient, computational papers are relatively rare.
Outhwaite et al. \cite{molero_jcsft_1992,outhwaite_jcsft_1993} compared the individual activity coefficients of primitive model electrolytes as obtained from various theories.
Similar studies have been published by S\o{}rensen et al. \cite{sorensen_jcsft_1989,sorensen_jcsft_1990,sorensen_jcsft_1991}, complemented by MC simulations.
Sloth \cite{sloth_cpl_1989-491} developed an expression for the single-ion activity coefficient on the basis of Kirkwood-Buff theory.
Ferse and M\"uller \cite{ferse_jsse_2011} factored the mean activity coefficient into individual ones using a product function optimized by experimental data.
Fraenkel \cite{fraenkel_jpcb_2012} applied his smaller-ion shell theory \cite{fraenkel_mp_2010} to estimate the individual activity coefficients of various electrolytes in comparison with the experimental data of VWV.

Lin and Lee \cite{lin_fpe_2003,lin_fpe_2005} used a parameterized equation with adjustable parameters. 
Their approach is similar to ours in the respect that they divided the excess chemical potential into terms corresponding to short- and long-range interactions.
The short-range part was intended to take solvation effects into account.
The parameterization of Lin and Lee, however, lacked the concentration-dependent dielectric constant. 
A similar approach was published by Pazuki and Rohani \cite{pazuki_fpe_2006} with different definitions for the two terms.
Taghikhani and Vera \cite{taghikhani_iecr_2000} correlated the experimental data with an MSA-based approach using concentration-dependent cation diameter.

Simulation studies impose the difficulties of system size effects due to violated charge neutrality when inserting individual ions (see Section \ref{subsec:exp}).
Sloth and S\o{}rensen simulated the individual activity coefficient since 1987 \cite{sloth_jcsft_1987,sloth_cpl_1988_140,sloth_cpl_1988_452,sorensen_jcsft_1989,sorensen_jcsft_1990,sorensen_jcsft_1991} using the Widom particle insertion method \cite{widom63} in the canonical ensemble.
They eventually developed a correction formula on the basis of a neutralizing background \cite{sloth_cpl_1990}. 
This correction term was used in the Adaptive Grand Canonical Monte Carlo (A-GCMC) method developed and used in our group \cite{malasics-jcp-132-244103-2010}.
Svensson and Woodward\ \cite{svensson_mp_1988} proposed a different correction method.

Lamperski and Pluciennik \cite{lamperski_ms_2009,lamperski_mp_2011} simulated various electrolyte models including the solvent primitive model (in that model, water molecules are represented by neutral hard spheres) using the GCMC algorithm developed by Lamperski \cite{lamperski_ms_2007}.  
Calculation of the individual activity coefficient is an especially hard problem for explicit water models because of the high density of the system.
We are aware of only a few works \cite{zhang_jctc_2010,joung_jcp_2013} that attempted the simulation of the individual activities for usual force fields using thermodynamic integration.

\begin{table}[t]
\renewcommand{\arraystretch}{1}
\begin{center}
\begin{tabular}{l|ll}
\hline
Salt & $\delta_{\mathrm{S}}$ & $b_{\mathrm{S}}$ \\
\hline
NaF & 15.45 & 3.76  \\ 
KF  & 12.4 & 2.2      \\ 
\hline
NaCl  & 15.45 & 3.76  \\ 
KCl   & 14.7 & 3.0  \\ 
RbCl  & 17.0 & 5.0  \\ 
CsCl  & 13.1 & 2.9  \\ 
\hline
LiBr  & 20.4 & 4.8  \\ 
NaBr  & 20.0 & 5.0  \\ 
KBr   & 14.6 & 2.5  \\ 
\hline
\end{tabular}  
\caption{Concentration dependence of the static dielectric constant of various pure electrolyte solutions. The table contains the $\delta_{\mathrm{S}}$ and $b_{\mathrm{S}}$ parameters of Eq.\ \ref{eq:epsc} for 1:1 electrolytes taken from Refs.\ \cite{fawcett-jpc-96,tikanen-jeac-97,buchner_jpca_1999}. For NaF, the data for NaCl were used. In the case of LiCl, the fit $\epsilon(c) = \epsilon_{\mathrm{w}} -15.5c +1.96c^{2} -0.306 c^{5/2} $ was used \cite{wei_rsi_1989}. For  2:1 electrolytes, the values $\delta_{\mathrm{S}}=34$ and $b_{\mathrm{S}}=10$ were used \cite{tikanen-bbgs-96}.
}
\label{tab:salts}
\end{center}
\end{table} 

\subsection{Experiments}
\label{subsec:exp}

Experimental approaches are based on measurements for an electrochemical cell composed of a reference electrode and a reversible electrode whose electrode potential depends on the activity of a single ion species.
The electrode that is selective for an ion species $i$ can be an ion selective membrane electrode (ISE) \cite{wilczekvera_fpe_1990,khoshkbarchi_aiche_1996,marcosarroyo1_jsc_996,rabie_jsc_1999,taghikhani_fpe_1999,taghikhani_cjce_2000,rodil_fpe_2001,rodil_aichej_2001,rodil_fpe_2003a,rodil_fpe_2003b,wilczekvera_aiche_2004,wilczekvera_fpe_2005,wilczekvera_fpe_2006b,wilczekvera_fpe_2006a,wilczekvera_ces_2007,wilczekvera_iecr_2009,wilczekvera_ces_2010,wilczekvera_fpe_2011,vera_jctd_442,wilczekvera_jctd_449,lee_jcice_2002,schneider_cit_2003,schneider_jsc_2004,zhuo_fpe_2008,sakaida_jpcb_2011}, an ion selective glass electrode \cite{shatkay_ac_1969,zhuo_fpe_2008}, or an electrode of the second type \cite{shatkay_ac_1969,hurlen_acssa_1979a,hurlen_acssa_1979b,hurlen_acssa_1981a}.
In the case of an electrode of the second type, Ag is immersed in a solution containing a $\mathrm{C} \mathrm{A}_{\nu_{-}}$ electrolyte and the sparingly soluble salt AgA as a precipitate.
This kind of electrode is selective for the anion.
The reference electrode (ref) is usually a saturated Ag/AgCl electrode.
The Electromotive Force of the cell is
\begin{equation}
E_{i} = E_{i,0} + S_{i} \ln \left(\gamma_{i} c_{i}\right) - E_{\mathrm{ref}}+ E_{\mathrm{J}}(c),
\label{eq:emf}
\end{equation}  
where $E_{i}^{0}$ is a standard potential that is constant in a given measurement, $E_{\mathrm{ref}}$ is the potential of the reference electrode, and $E_{\mathrm{J}}(c)$ is the junction potential raised at the interface of the two solutions.
The junction potential generally depends on the electrolyte concentration, although efforts to develop a salt bridge with a stable junction potential have been made \cite{sakaida_jpcb_2011}.
The common drawback of all these measurements is that the junction potential cannot be measured directly, so its determination requires some theoretical consideration. 
They commonly calculate it from Henderson's equation \cite{henderson_zpc_1907,henderson_zpc_1908} or any of its modifications \cite{wilczekvera_aiche_2004,harned_jpc_1926,bates1964determination,bates1973determination,harper_jpc_1985}. 

The proposal of the VWV group is that $\gamma_{i}$ in this equation can be identified with the equilibrium activity coefficient in the case of an ISE.
They claim that if the measurement is done continuously in a limited amount of time (a few hours), while increasing the concentration from infinite dilution towards saturation, the parameters $E_{i}^{0}$ and $S_{i}$ do not change during the experiment.
This seems to a be a crucial point in their arguments.
We do not discuss the debate between the VWV group and those who question their method (Malatesta \cite{malatesta_jsc_2000,malatesta_fpe_2005,malatesta_fpe_2006,malatesta_fpe_2010} and Zarubin \cite{zarubin_jctd_2011,zarubin_jctd_445,zarubin_jctd_451}); we direct the reader to the original papers instead. \cite{wilczekvera_fpe_2005,wilczekvera_fpe_2006b,wilczekvera_ces_2010,wilczekvera_fpe_2011,vera_jctd_442,wilczekvera_jctd_449,malatesta_jsc_2000,malatesta_fpe_2005,malatesta_fpe_2006,malatesta_fpe_2010,malatest_ces_2010,zarubin_jctd_2011,zarubin_jctd_445,zarubin_jctd_451,zarubin_jpcb_2012_comment,fraenkel_jpcb_2012_reply}.

At the heart of the debate, however, there seems to be a statement about which we have a definite opinion.
This statement originating from Guggenheim \cite{guggenheim_jpc_1929,guggenheim_jpc_1930,guggenheim_jpc_1930b} and Taylor \cite{taylor_jpc_1927} is that activity has ``no physical significance for a single ion species''.
The root of this opinion, in turn, is that one cannot add a measurable quantity of ions to the system without also adding the same amount of counterions; otherwise, one would violate the condition of charge neutrality.
Our opinion, on the other hand, is in accordance with that of Lewis \cite{lewis1923}, Harned \cite{harned_jpc_1926}, and Br\o{}nsted \cite{bronsted_zpc_1929,bronsted_zpc_1932}: the individual activity is a thermodynamically well-defined quantity.
The requirement of charge neutrality is a concept of the macroscopic world and an effect of long-time averages. 
Spontaneous violations of charge neutrality instantaneously and locally, however, are common and expected.

This is especially true in a simulation that is supposed to mimic reality on the microscopic level.
One can add one single ion to an electrolyte momentarily without a problem.
Violation of charge neutrality is ``punished'' by a lowered acceptance probability of this insertion. 
In a next simulation step (either time step or MC step), charge imbalance is likely to be corrected by deletion of the extra ion.
In this way, charge fluctuates in an open system, but it fluctuates around zero thus producing a charge neutral system on average.

The issue can be lighted through Widom's particle insertion method \cite{widom63}.
In this simulation technique, a test particle is inserted into the system randomly, the energy cost of the insertion, $\Delta U$, is computed and the excess chemical potential is obtained as
\begin{equation}
\mu^{\mathrm{EX,Widom}}_{i}=kT\ln \left\langle \exp \left(-\dfrac{\Delta U}{kT} \right) \right\rangle,
\end{equation} 
where the brackets denote an ensemble average over numerous such insertions.
If we insert an ion, the interaction with the missing counterion is absent, so the equation converges to the correct result only in the limit of an infinitely large system.
Simulations, on the other hand, necessarily use a finite simulation cell, so the finite system size error is always present. 
Sloth and S\o{}rensen \cite{sloth_cpl_1990} suggested a correction term to estimate the error.
They assumed that the missing charge, $Q$, is smeared over the cubic simulation cell ($V=L^{3}$) as a constant volume charge of magnitude $Q/V$.
The interaction of an inserted ion, $q$, with this neutralizing background can be integrated as
\begin{equation}
\mu^{\mathrm{corr}}_{Q} = -\dfrac{qQ}{32\pi \epsilon_{0}\epsilon L} K ,
\label{eq:screening}
\end{equation} 
where $K$ is a constant (see Ref.\ \cite{sloth_cpl_1990}).
When we insert a single ion in a charge neutral solution, $Q=-q$.
The correction term scales with $L^{-1}$, so it goes to zero as the size of the system approaches infinity.

The chemical potential of a charged species, therefore, can unambiguously be defined as a partial molar quantity in the thermodynamic limit:
\begin{equation}
\mu_{i}=\lim_{V\rightarrow\infty} \left( \dfrac{\partial F}{\partial n_{i}} \right)_{T,V,n_{j}\neq n_{i}} .
\end{equation} 
Although this looks clear fundamentally, developing an actual method with which $\mu_{i}$ can be measured is far from being trivial.
However, condemning the concept of the chemical potential of an ion as senseless just because it is hard to design an appropriate experiment is putting the cart before the horse.
Therefore, we accept the works of experimentalists \cite{wilczekvera_fpe_1990,khoshkbarchi_aiche_1996,marcosarroyo1_jsc_996,rabie_jsc_1999,taghikhani_fpe_1999,taghikhani_cjce_2000,rodil_fpe_2001,rodil_aichej_2001,rodil_fpe_2003a,rodil_fpe_2003b,wilczekvera_aiche_2004,wilczekvera_fpe_2005,wilczekvera_fpe_2006b,wilczekvera_fpe_2006a,wilczekvera_ces_2007,wilczekvera_iecr_2009,wilczekvera_ces_2010,wilczekvera_fpe_2011,vera_jctd_442,wilczekvera_jctd_449,shatkay_ac_1969,hurlen_acssa_1979a,hurlen_acssa_1979b,hurlen_acssa_1981a,lee_jcice_2002,schneider_cit_2003,schneider_jsc_2004,zhuo_fpe_2008,sakaida_jpcb_2011} as honest efforts to determine this well-defined physico-chemical quantity from reproducible experiments.
We will use the data of the VWV group \cite{wilczekvera_fpe_1990,khoshkbarchi_aiche_1996,marcosarroyo1_jsc_996,rabie_jsc_1999,taghikhani_fpe_1999,taghikhani_cjce_2000,rodil_fpe_2001,rodil_aichej_2001,rodil_fpe_2003a,rodil_fpe_2003b,wilczekvera_aiche_2004,wilczekvera_fpe_2005,wilczekvera_fpe_2006b,wilczekvera_fpe_2006a,wilczekvera_ces_2007,wilczekvera_iecr_2009,wilczekvera_ces_2010,wilczekvera_fpe_2011,vera_jctd_442,wilczekvera_jctd_449} and Hurlen \cite{hurlen_acssa_1979a,hurlen_acssa_1979b,hurlen_acssa_1981a} for comparison, but we will also discuss other experiments in Section \ref{sec:results}.

\section{Method}
\label{sec:method}

In this section, we present the methods with which we calculate the II and IW parts separately (see Eq.\ \ref{eq:ii+iw}).
We emphasize that the methods and the models behind them (Eq.\ \ref{eq:pm}) are not unique; different ion models and different approaches for solvation can be chosen. 
Although the two terms can be computed independently, there is one crucial quantity that connects them: the experimental concentration-dependent dielectric constant.
The concentration dependence has been taken from various sources \cite{fawcett-jpc-96,tikanen-bbgs-96,tikanen-jeac-97,buchner_jpca_1999,wei_rsi_1989} and fitted with the following equation
\begin{equation}
\epsilon(c)=\epsilon_{\mathrm{w}} - \delta_{\mathrm{S}} c + b_{\mathrm{S}} c^{3/2},
\label{eq:epsc}
\end{equation} 
where $\epsilon_{\mathrm{w}}=78.45$ is the dielectric constant of water (the infinitely dilute solution) at $K=298.15$ K.
The coefficients of the equation for the electrolytes studied in this work are found in Table \ref{tab:salts} and in the caption of the table.

\subsection{Calculation of the II term}
\label{sec:calc-II}

The II term is calculated on the basis of the PM of electrolytes using the A-GCMC simulation method of Malasics and Boda \cite{malasics-jcp-132-244103-2010,malasics-jcp-128-124102-2008}.
This procedure works in the grand canonical ensemble, where the chemical potential is the independent variable instead of the concentration. 
Determination of the chemical potentials, $\mu_{i}$, that correspond to prescribed (targeted) concentrations, $c_{i}^{\mathrm{targ}}$, therefore, requires an iterative procedure.
The chemical potential for the $(n+1)$th iteration is estimated from
\begin{equation}
 \mu_{i}(n+1) = \mu_{i}(n) + kT\ln \dfrac{c_{i}^{\mathrm{targ}}}{\left\langle c_{i}(n)\right\rangle} - \dfrac{z_{i}e\left\langle Q(n) \right\rangle}{32\pi \epsilon_{0}\epsilon L}K  ,
\label{eq:hegesztett}
\end{equation} 
where $\mu_{i}(n)$ is the chemical potential of species $i$ in the $n$th iteration, $\left\langle c_{i}(n)\right\rangle$ is the concentration obtained from a GCMC simulation in the $n$th iteration, while the last term is the correction term corresponding to the average net charge of the cubic ($V=L^{3}$) simulation cell in the $n$th iteration, $\left\langle Q(n) \right\rangle$ (see Eq.\ \ref{eq:screening}).  
The algorithm of the A-GCMC method is robust and converges to the desired values fast.
After convergence, the chemical potentials fluctuate around the correct values; the final results, therefore, are obtained from running averages.

The underlying model requires establishing the molecular parameters, $R_{i}$ and $z_{i}$, and the thermodynamic parameter, $\epsilon(c)$.
We emphasize that the radii of the ``bare ions'' (independent of $c$) are used for $R_{i}$ (the Pauling radii, Table \ref{tab:ions}) and experimentally measured values  are used for $\epsilon(c)$ (Table \ref{tab:salts}).

\subsection{Calculation of the IW term}
\label{sec:iw}

The IW term is obtained from experimental data using a parameterization based on Born's treatment of solvation \cite{born}.
In this theory, the solvation free energy, $\Delta G_{i}^{\mathrm{s}}$, is assumed to be equal to the electrostatic energy change of the inversion of a spherical ion of radius $R_{i}^{\mathrm{B}}$ in the continuum of dielectric constant $\epsilon(c)$ and is given as 
\begin{equation}
 \Delta G_{i}^{\mathrm{s}} (c) = \dfrac{z_{i}^{2}e^{2}}{8\pi \epsilon_{0} R^{\mathrm{B}}_{i}} \left( \dfrac{1}{\epsilon (c)} -1\right) .
\label{eq:born}
\end{equation} 
The IW part of the excess chemical potential is defined as the difference in the solvation free energy of the concentrated and dilute solutions: 
\begin{equation}
\mu_{i}^{\mathrm{IW}} (c) = \Delta G_{i}^{\mathrm{s}}(c) - \Delta G_{i}^{\mathrm{s}} =
\dfrac{z_{i}^{2}e^{2}}{8\pi\epsilon_{0} R^{\mathrm{B}}_{i}} \left( \dfrac{1}{\epsilon (c)} -\dfrac{1}{\epsilon_{\mathrm{w}}} \right) ,
\label{eq:deltaU}
\end{equation} 
where $\Delta G_{i}^{\mathrm{s}}=\Delta G_{i}^{\mathrm{s}}(c\rightarrow 0)$ is the experimental solvation (hydration) energy in water at temperature $T=298.15$ K (Table \ref{tab:ions}).
It is important to note that the radius $R_{i}^{\mathrm{B}}$ (the Born radius, Table \ref{tab:ions}) does not have to be the same as $R_{i}$ used in the calculation of the II term.
It is obtained from Eq.\ \ref{eq:born} by writing it up for the case of infinitely dilute electrolyte:
\begin{equation}
 \Delta G_{i}^{\mathrm{s}} = \dfrac{z_{i}^{2}e^{2}}{8\pi \epsilon_{0} R^{\mathrm{B}}_{i}} \left( \dfrac{1}{\epsilon_{\mathrm{w}}} -1\right) .
\label{eq:bornw}
\end{equation}
Expressing $R_{i}^{\mathrm{B}}$ from Eq.\ \ref{eq:bornw} and substituting it into Eq.\ \ref{eq:deltaU}, we obtain an expression for the IW term that contains only experimental parameters:
\begin{equation}
 \mu_{i}^{\mathrm{IW}}(c) = \Delta G_{i}^{\mathrm{s}} \dfrac{\epsilon(c)-\epsilon_{\mathrm{w}}}{\epsilon(c) \; (\epsilon_{\mathrm{w}}-1)}.
\label{eq:iwscaled}
\end{equation} 
This equation describes the $\epsilon (c)$-dependence of the IW term and gives the correct hydration free energy in the $c\rightarrow 0$ limit.
Note that $\mu_{i}^{\mathrm{IW}}(c)>0$ because $\epsilon(c)<\epsilon_{\mathrm{w}}$ and $\Delta G_{i}^{\mathrm{s}}<0$.

\section{Results}
\label{sec:results}

\begin{figure}[t]
\begin{center}
\includegraphics*[width=0.85\linewidth]{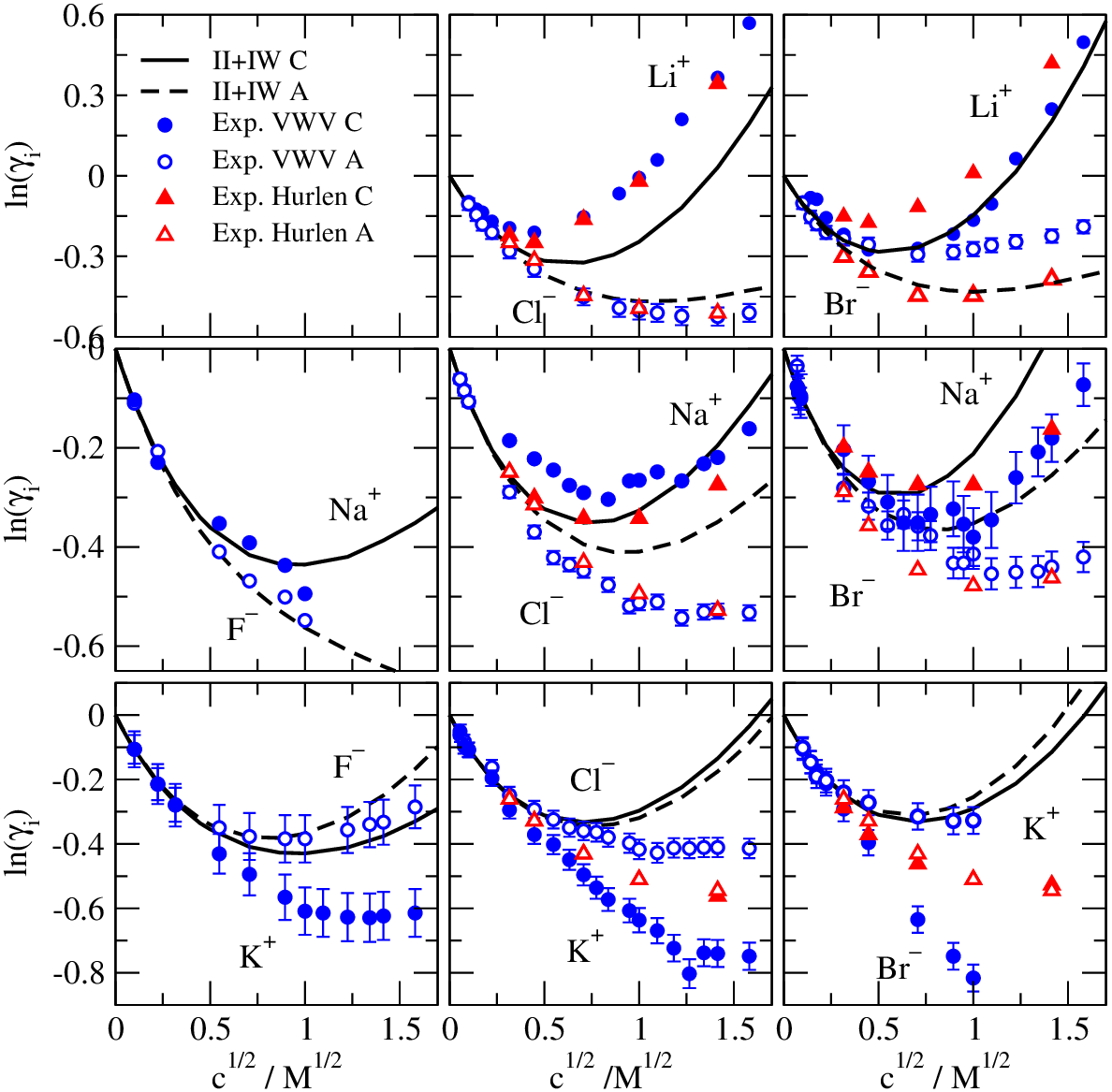}
\end{center}
\caption{Individual activity coefficients for 1:1 electrolytes. 
Solid and dashed lines refer to the II+IW results for cation (C) and anion (A), respectively.
Filled and open symbols refer to experimental data for cation and anion, respectively.
Blue circles are the data of the VWV group \cite{wilczekvera_aiche_2004}, while red triangles are the data of Hurlen \cite{hurlen_acssa_1979a,hurlen_acssa_1981a}.
Error bars for experimental data are shown in some representative cases.
}
\label{fig1}
\end{figure}

\begin{figure}[t]
\begin{center}
\includegraphics*[width=0.65\linewidth]{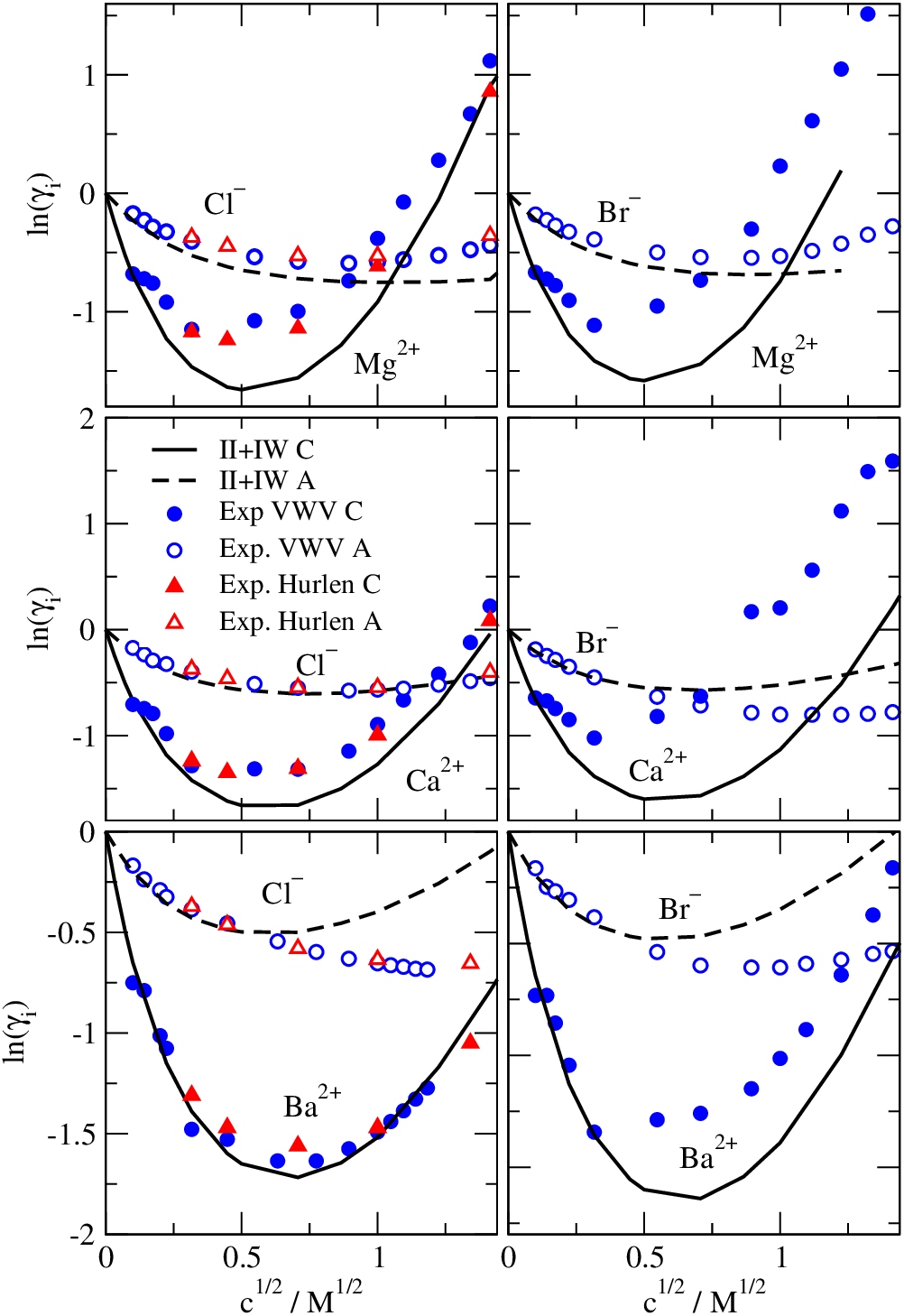}
\end{center}
\caption{Individual activity coefficients for 2:1 electrolytes.
Solid and dashed lines refer to the II+IW results for cation (C) and anion (A), respectively.
Filled and open symbols refer to experimental data for cation and anion, respectively.
Blue circles are the data of the VWV group \cite{wilczekvera_aiche_2004}, while red triangles are the data of Hurlen \cite{hurlen_acssa_1979b}.
}
\label{fig2}
\end{figure}

In this paper, we report results for 1:1 electrolytes LiCl, LiBr, NaF, NaCl, NaBr, KF, KCl, and KBr, and for 2:1 electrolytes MgCl$_{2}$, MgBr$_{2}$, CaCl$_{2}$, CaBr$_{2}$, BaCl$_{2}$, and BaBr$_{2}$.
Individual activity coefficients are shown in Fig.\ \ref{fig1} for the 1:1 systems and in Fig.\ \ref{fig2} for the 2:1 systems.
The results obtained from the II+IW model are shown in comparison with the experimental data of the VWV group \cite{wilczekvera_aiche_2004} and Hurlen \cite{hurlen_acssa_1979a,hurlen_acssa_1979b,hurlen_acssa_1981a}. 
The agreement between the two experiments is usually quite good (much better in the 2:1 case) despite the fact that the two experimental setups used different working electrodes; ISE by the VWV group \cite{wilczekvera_aiche_2004} and a reversible anion specific electrode of the second kind by Hurlen \cite{hurlen_acssa_1979a,hurlen_acssa_1979b,hurlen_acssa_1981a} (he measured only for the anions; the cation value was computed from that and the mean).
The agreement indicates the reliability of the overall procedure (use of ion specific electrodes), while disagreement (especially for potassium halides) sheds light on the possible weaknesses of the electrochemical method.

A general quantitative observation is that the experimental $\ln \gamma_{i}$ changes in a more narrow range in the 1:1 case (between -0.8 and 0.6) than in the 2:1 case (between -1.6 and 1.6). 
Our theoretical curves reproduce this behavior by and large with different accuracy for different electrolytes as discussed below.

The qualitative behavior of the $\ln \gamma_{i}(c)$ curves is also different in the two cases. 
In the 1:1 case, $\ln \gamma_{i}$ is larger for one of the two species than for the other one over the entire concentration range ($\Delta \mu^{\mathrm{EX}}$ does not change sign).
In the 2:1 case, on the other hand, $\ln \gamma_{i}$ is larger for the anion at smaller concentrations, while it is larger for the cation at larger concentrations in most cases ($\Delta \mu^{\mathrm{EX}}$ changes sign).
A very important result is that this behavior is reproduced by our model (see Fig.\ \ref{fig2}).

The performance of our theory can be judged from various aspects.
We can look at the individual $\gamma_{i}$ curves and check the agreement with the experimental data.
The other way is to check the agreement for the mean, $\mu_{\pm}^{\mathrm{EX}}$, and the difference, $\Delta\mu^{\mathrm{EX}}$.
\textbf{The question then arises that how the accuracy of our model correlates for these quantities.}
If the model is good for the mean, does it mean that it is also good for the difference?
Or vice versa, if the model fails for the mean, can its prediction be still fine for the difference?
This latter happens when we introduce the same error in the $\mu_{i}^{\mathrm{EX}}$ of the cation and the anion.
This scenario is surely possible.
The opposite (that we introduce errors of opposite signs), however, is also possible. 

The behavior of our model for the mean was discussed in our previous papers in detail \cite{vincze-jcp-133-154507-2010,valisko-jcp-140-234508-2014}.
The general conclusion was that the II+IW model works surprisingly well for 2:1 electrolytes, while it has problems in the 1:1 case.
Specifically, the model cannot provide good $\gamma_{\pm}$ data for large cations (K$^{+}$, Cs$^{+}$).
Also, the dependence of $\gamma_{\pm}$ on the cation radius shows the opposite behavior in the model and in experiment (Fig.\ 2 of Ref.\ \cite{vincze-jcp-133-154507-2010}).
The results of Figs.\ \ref{fig1} and \ref{fig2} confirm these findings, but now they provide deeper insight into the mechanisms because they show individual activity coefficients. 

For LiCl and LiBr, our results agree well with the experimental data for the anion, while they underestimate $\ln\gamma_{i}$ for Li$^{+}$. 
The agreement between the experimental data of VWV and Hurlen is good for LiCl, but quite bad for LiBr.

In the case of sodium halides, we rather overestimate the experimental data, although the agreement for NaF is quite good up to the concentration (1 M) for which experimental data are available.
More experimental data are available for NaCl beyond those shown in Fig.\ \ref{fig1}.
NaCl was also studied by Zhuo et al. \cite{zhuo_fpe_2008} (good agreement with the VWV data), Lee  et al. \cite{lee_jcice_2002} (they overestimate the difference, $\Delta\mu^{\mathrm{EX}}$), and Shatkay and Lerman \cite{shatkay_ac_1969} (they underestimate both $\gamma_{i}$'s).
For NaCl, a detailed analyzis on the sensitivity of our theory on the model parameters has been published \cite{valisko-jcp-140-234508-2014}. 
For NaBr, interestingly, the data of VWV agree with those of Hurlen, but both are different from other measurements (Lee et al.\ \cite{zhuo_fpe_2008} and Zhuo et al.\ \cite{lee_jcice_2002}; they agree with each other well).
 
Although our model does not work properly for potassium halides, this is a case that we find quite interesting.
Hurlen \cite{hurlen_acssa_1979a,hurlen_acssa_1979b,hurlen_acssa_1981a} deduced from his measurements that $\gamma_{i}$ is practically the same for K$^{+}$ and Cl$^{-}$ ($\Delta\mu^{\mathrm{EX}}\approx 0$).
The VWV group \cite{wilczekvera_aiche_2004}, on the other hand, predicts that $\gamma_{i}$ is smaller for K$^{+}$ than for Cl$^{-}$ (note that the work of Dash et al. \cite{dash_isrn_2012} predicts the opposite trend).
Although our results for the mean are really off, our data for the difference support Hurlen's findings (see Fig.\ \ref{fig1}). 
We will discuss this case in the Discussions (Sec.\ \ref{sec:discuss}) in more detail. 

Electrolytes, with a divalent cation (2:1 systems, see Fig.\ \ref{fig2}) provide a much better case study for our theory than 1:1 electrolytes do.
The two ions have different charges in the 2:1 case.
Therefore, their interactions with their ionic environment are very different, so the II term is very different for the two ions.
The hydration free energies also differ considerably; they are much larger for divalents in absolute value (see Table \ref{tab:ions}).
Consequently, the IW term is also very different for the two ions.
There are strong asymmetries between the two ions resulting in the anomalous behavior seen in Fig.\ \ref{fig2}.

By ``anomalous'' we mean that the activity of the cation is smaller than that of the anion for small concentrations, while the reverse is true for larger concentrations. 
In other words, the difference $\Delta \mu^{\mathrm{EX}}(c)$ has a minimum and changes sign.
This behavior is absent in the case of 1:1 electrolytes because the charge asymmetry of the ions is absent.

Also, this anomalous effect is powerful in the sense that $\ln \gamma_{i}$ changes over a wide range (in $kT$ units).
This indicates that the nonmonotonic behavior is the result of the balance of large energy terms. 
Theories dealing with this phenomenon should account for these large energy terms.
As Fig.\ \ref{fig2} shows, our theory reproduces this phenomenon qualitatively.
This implies that the two basic (free) energetic terms introduced in our treatment (the II and IW terms) contain sufficient information to provide qualitative description.
For quantitative agreement, of course, further microscopic information and more detailed models would be necessary. 
Detailed discussion about how these terms work together to provide the necessary balance is given in the next section.

\section{Discussion}
\label{sec:discuss}

After presenting the raw data for the individual activity coefficients, we turn to understanding our results better by considering the II and IW terms separately.
Because we have too much data in Figs.\ \ref{fig1} and \ref{fig2}, we choose two representative cases to discuss the details, NaCl and CaCl$_{2}$.

\begin{figure}[t]
\begin{center}
\includegraphics*[width=0.6\linewidth]{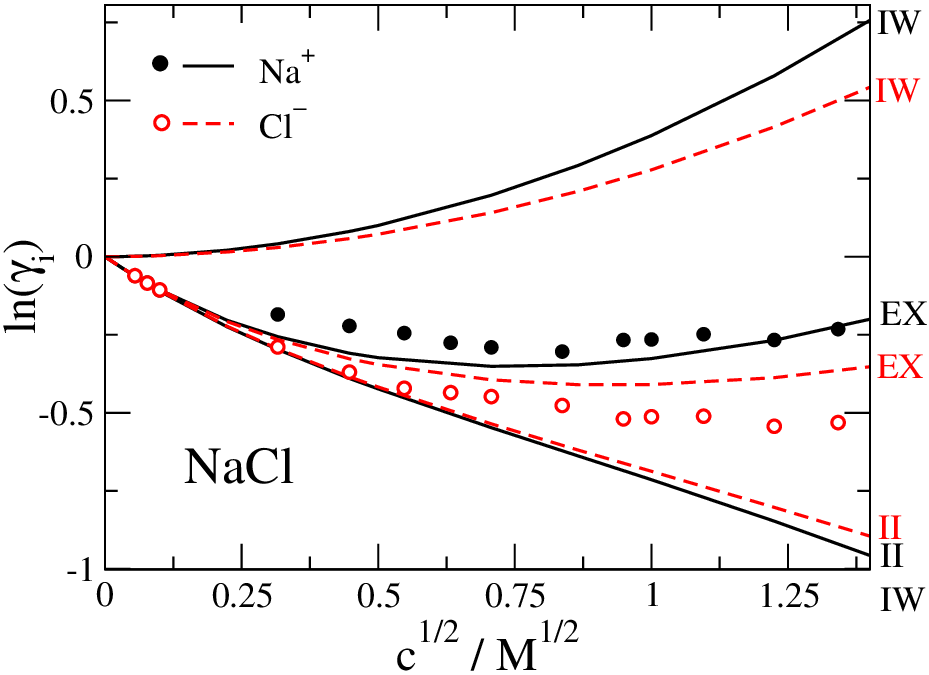}
\end{center}
\caption{II and IW components of $\ln \gamma_{i}$ (EX) for Na$^{+}$ and Cl$^{-}$ in NaCl electrolytes as functions of concentration.
Solid black and dashed red curves refer to Na$^{+}$ and Cl$^{-}$, respectively.
Filled and open circles are the experimental data of the VWV group \cite{wilczekvera_aiche_2004} for Na$^{+}$ and Cl$^{-}$, respectively.}
\label{fig3}
\end{figure}

Figure \ref{fig3} shows the results for NaCl.
The II values are slightly larger (in magnitude) for Na$^{+}$ because Na$^{+}$ has stronger interaction with the surrounding ions due to its smaller size.
Similarly, the IW values are slightly larger for Na$^{+}$ because Na$^{+}$ has stronger interaction with the surrounding water, also due to its smaller size.
In our treatment, this stronger IW interaction is because the experimental solvation free energy is larger (in magnitude) for Na$^{+}$ (see Table \ref{tab:ions}).
The sum of the II and IW curves (the EX curves in Fig.\ \ref{fig3}) is smaller for Cl$^{-}$ than for Na$^{+}$, in line with experiments.
This means that their difference, the $\Delta \mu^{\mathrm{EX}}$ term (see Eq.\ \ref{eq:delta_mu}), is positive. 
The IW term has a dominant effect in this, as discussed later.

\begin{figure}[t]
\begin{center}
\includegraphics*[width=0.6\linewidth]{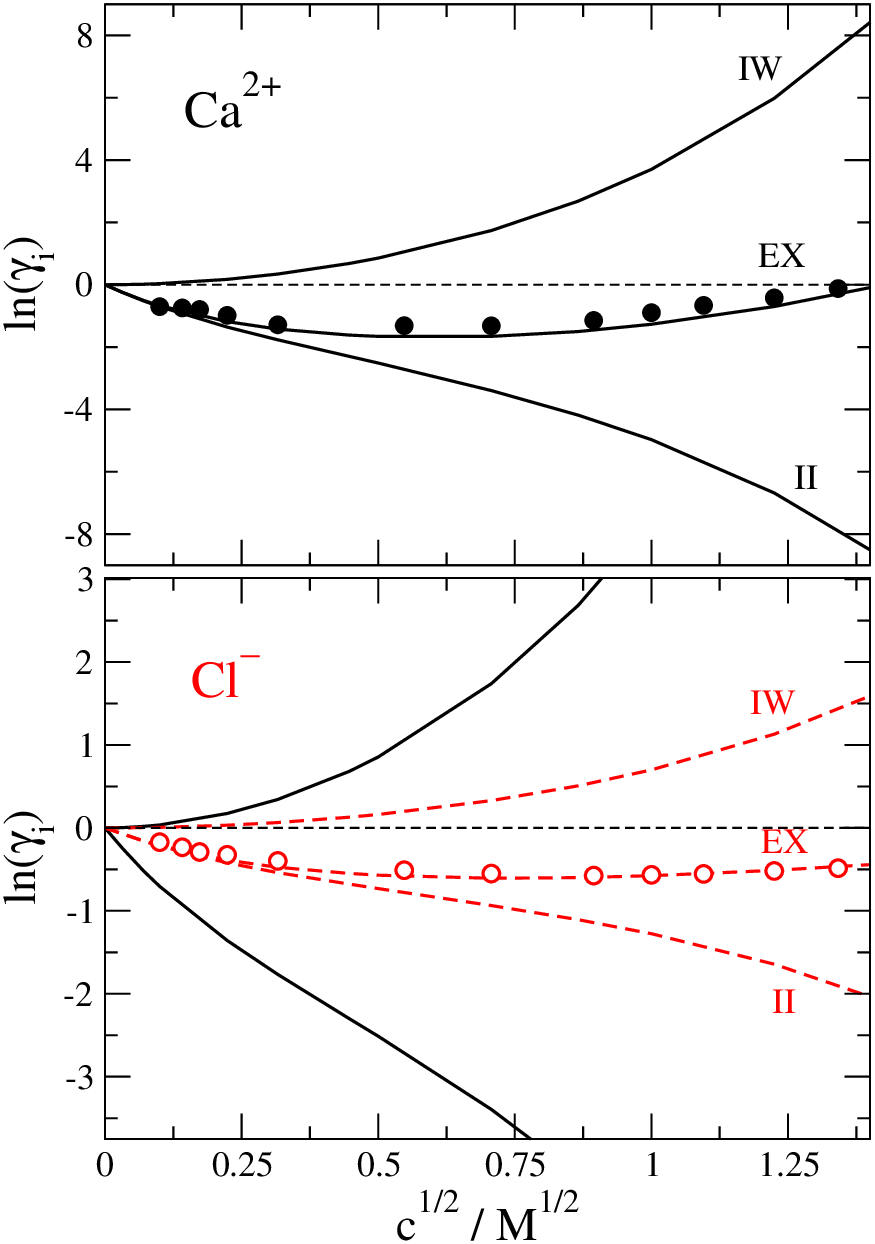}
\end{center}
\caption{II and IW components of $\ln \gamma_{i}$ (EX) for Ca$^{2+}$ (top panel) and Cl$^{-}$ (bottom panel) in CaCl$_{2}$ electrolytes  as functions of concentration.
Solid black and dashed red curves refer to Ca$^{2+}$ and Cl$^{-}$, respectively.
Filled and open circles are the experimental data of the VWV group \cite{wilczekvera_aiche_2004} for Ca$^{2+}$ and Cl$^{-}$, respectively.
The II and IW curves for Ca$^{2+}$ are also shown in the bottom panel for comparison (solid lines).}
\label{fig4}
\end{figure}

Figure \ref{fig4} shows the results for CaCl$_{2}$.
Here, the top panel shows the data for Ca$^{2+}$, while the bottom panel for Cl$^{-}$.
The bottom panel also shows the II and IW curves for Ca$^{2+}$ to emphasize the difference in the scales of the ordinates of the two panels.
In particular, the II and IW values range between $-8.5kT$ and $8kT$ for Ca$^{2+}$, while only between $-2kT$ and $1.5kT$ for Cl$^{-}$.
The large variation in the case of Ca$^{2+}$ results in an EX curve that has a deeper minimum and steeper upswing at larger concentrations (also, see Fig.\ \ref{fig2}).
The more shallow EX curve for Cl$^{-}$ is a result of the balance of the smaller (in magnitude) II and IW terms.

Now let us turn our attention to discussing the difference of the excess chemical potentials for the cation and the anion, $\Delta\mu^{\mathrm{EX}}$ (see Eq.\ \ref{eq:delta_mu}).
Similar to the EX term, we can also define the differences for the II and IW terms, $\Delta\mu^{\mathrm{II}}=\mu^{\mathrm{II}}_{+}-\mu^{\mathrm{II}}_{-}$ and $\Delta\mu^{\mathrm{IW}}=\mu^{\mathrm{IW}}_{+}-\mu^{\mathrm{IW}}_{-}$.

\begin{figure}[t]
\begin{center}
\includegraphics*[width=0.6\linewidth]{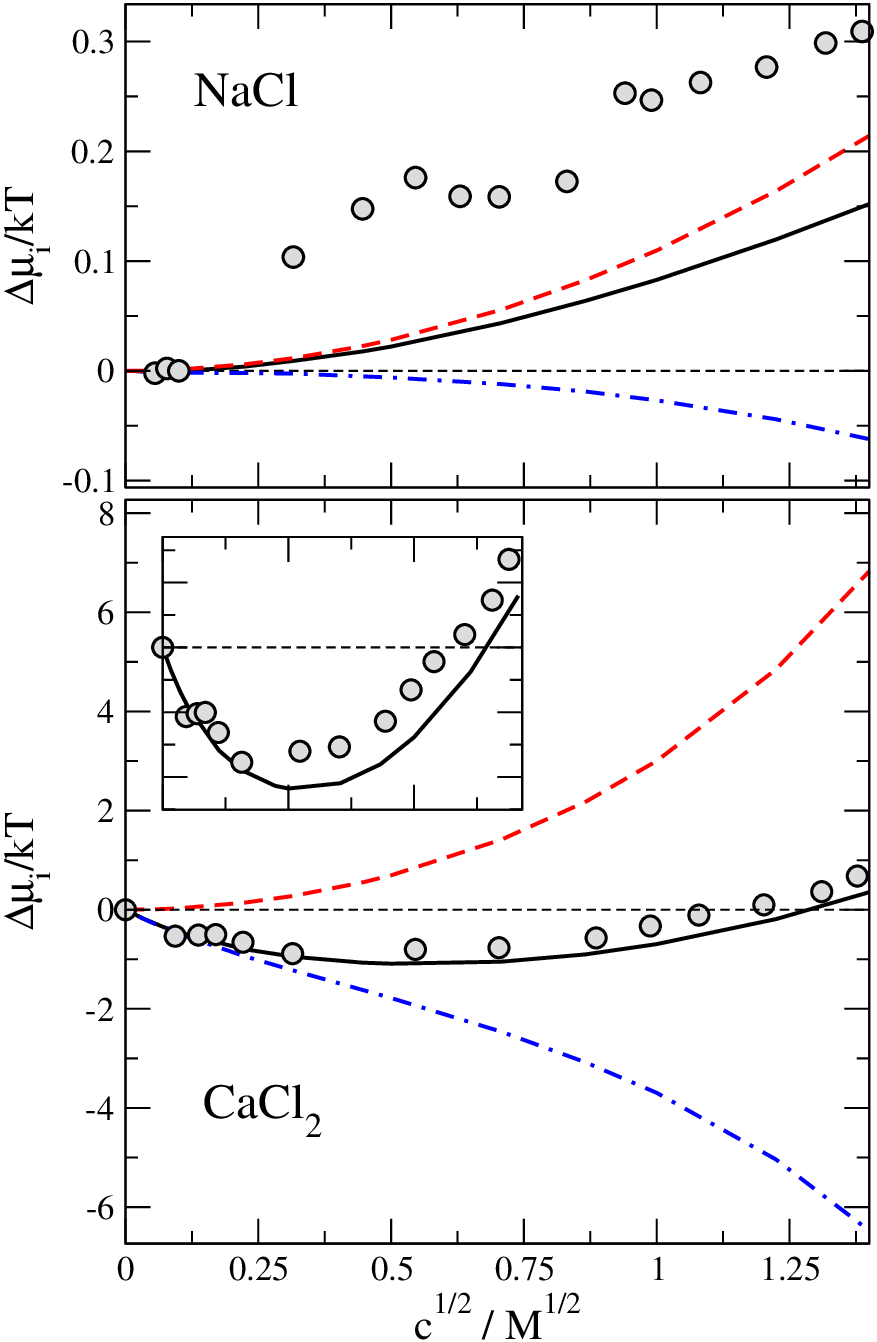}
\end{center}
\caption{The differences of the excess chemical potential components of cations and anions (see Eq.\ \ref{eq:delta_mu}) as functions of concentrations. 
Solid, dot-dashed, and dashed lines refer to the $\Delta \mu^{\mathrm{EX}}(c)$, $\Delta \mu^{\mathrm{II}}(c)$, and $\Delta \mu^{\mathrm{IW}}(c)$ curves, respectively. 
The top and bottom panels show the data for CaCl$_{2}$ and NaCl electrolytes, respectively.
Symbols are the experimental data of the VWV group \cite{wilczekvera_aiche_2004}.
The inset in the top panel shows the $\Delta \mu^{\mathrm{EX}}(c)$ data for CaCl$_{2}$.}
\label{fig5}
\end{figure}

The top panel of Fig.\ \ref{fig5} shows these curves for NaCl.
As already seen at Fig.\ \ref{fig3}, the $\Delta\mu^{\mathrm{EX}}$ term is always positive.
The dominant term that determines the sign of $\Delta\mu^{\mathrm{EX}}$ is the IW term. 
Without the IW term, the II term alone would provide a negative EX term in contrast to experiments, at least, using the Pauling diameter.
To achieve a large positive $\Delta\mu^{\mathrm{EX}}$ value without the IW term (using concentration independent dielectric constant), we would need larger ionic radius for Na$^{+}$ than for Cl$^{-}$.
Our reasoning against such fitted ``solvated'' radii has been given elsewhere \cite{vincze-jcp-133-154507-2010,vincze-jcp-134-157102-2011,valisko-jcp-140-234508-2014}.

For CaCl$_{2}$, on the other hand (bottom panel of Fig.\ \ref{fig5}), the balance of the IW and II terms is more complex.
For low concentrations, the II term dominates, so the EX curve decreases together with the II curve in accordance with the DH theory.
At larger concentrations, on the other hand, the IW term becomes more and more dominant so that it tips the balance in favor of the IW term above a certain concentration (the $\Delta\mu^{\mathrm{EX}}$ curve changes sign, see inset).
This kind of balance of the II and IW terms and their different $c$-dependence were also the reasons for the success of the II+IW theory in reproducing the nonmonotonic behavior of the mean excess chemical potential \cite{vincze-jcp-133-154507-2010,vincze-jcp-134-157102-2011,valisko-jcp-140-234508-2014}.

Our results show that the two simple theories that we use to estimate the II and IW terms contain all the necessary physics to provide this behavior.
The II term is computed on the basis of the charged hard sphere model (Eq.\ \ref{eq:pm}), where these ions are swimming in a dielectric background characterized by the concentration-dependent dielectric constant, $\epsilon (c)$. 
In this term, the determining factors are the ionic charges and radii (besides $\epsilon(c)$, of course).
The IW term is computed from a parameterization of the interaction of an ion with the surrounding dielectric, where the parameterization is based on two experimentally measurable quantities, the dielectric constant and the hydration free energy (Eq.\ \ref{eq:iwscaled}). 
The functional form of the IW term depending on these quantities is obtained from the simplest possible solvation theory we can think of, Born's treatment.
In this term, the determining factor is the hydration free energy (besides $\epsilon(c)$, of course).

\begin{figure}[t]
\begin{center}
\includegraphics*[width=0.6\linewidth]{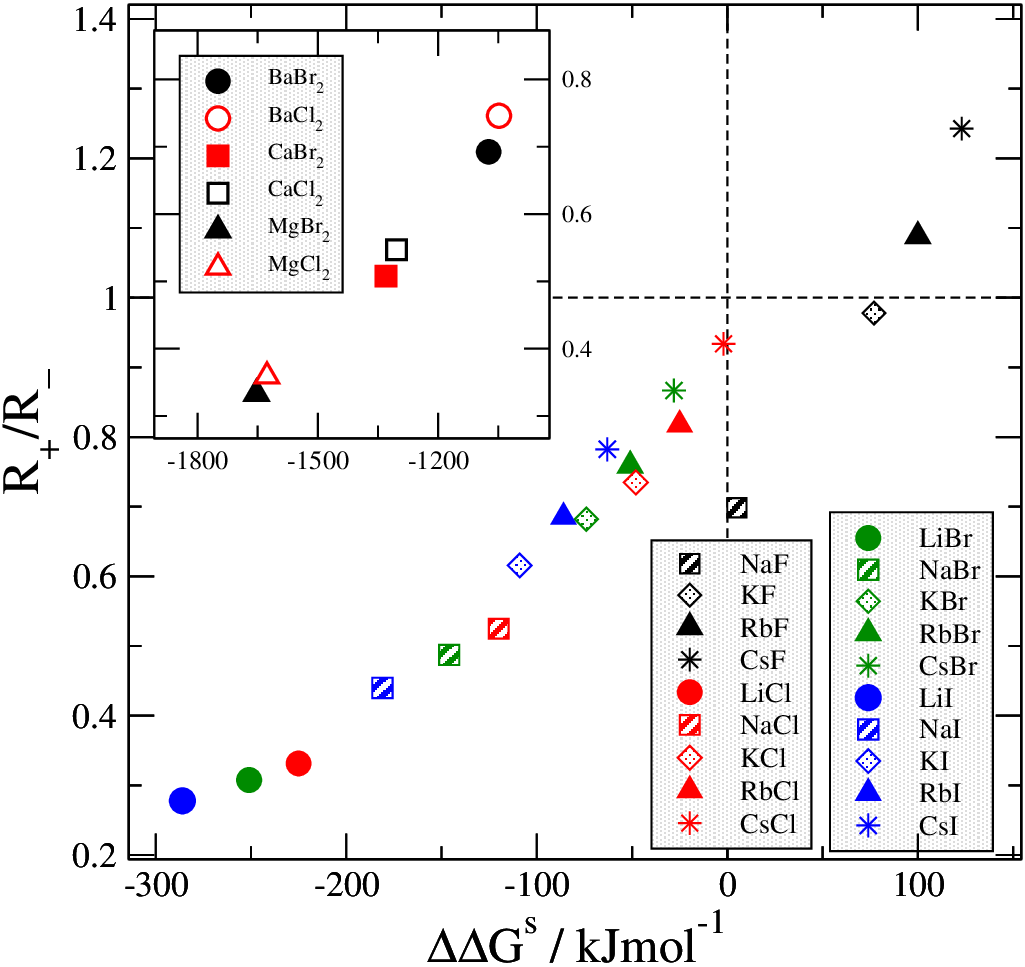}
\end{center}
\caption{The ratio of the Pauling radii ($R_{+}/R_{-}$) \cite{pauling} vs.\ the difference of the experimental hydration free energies ($\Delta\Delta G^{\mathrm{s}}=\Delta G^{\mathrm{s}}_{+}-\Delta G_{-}^{\mathrm{s}}$) \cite{fawcett-book} for various electrolytes.}
\label{fig6}
\end{figure}

If we fix the $\epsilon(c)$ function, for 1:1 electrolytes, the two determining factors are the ionic radii and the hydration free energies.
The difference $\Delta\mu^{\mathrm{EX}}$, therefore, depends on the difference of the hydration free energies for the cation and the anion, $\Delta\Delta G^{\mathrm{s}}=\Delta G_{+}^{\mathrm{s}}-\Delta G_{-}^{\mathrm{s}}$, as well as on the ratio of the cation/anion radii, $R_{+}/R_{-}$.
These quantities strongly correlate, as shown in Fig.\ \ref{fig6}.
This fact supports the Born formalism, and, indirectly, our approach to handle the IW term.

\begin{figure}[t]
\begin{center}
\includegraphics*[width=0.6\linewidth]{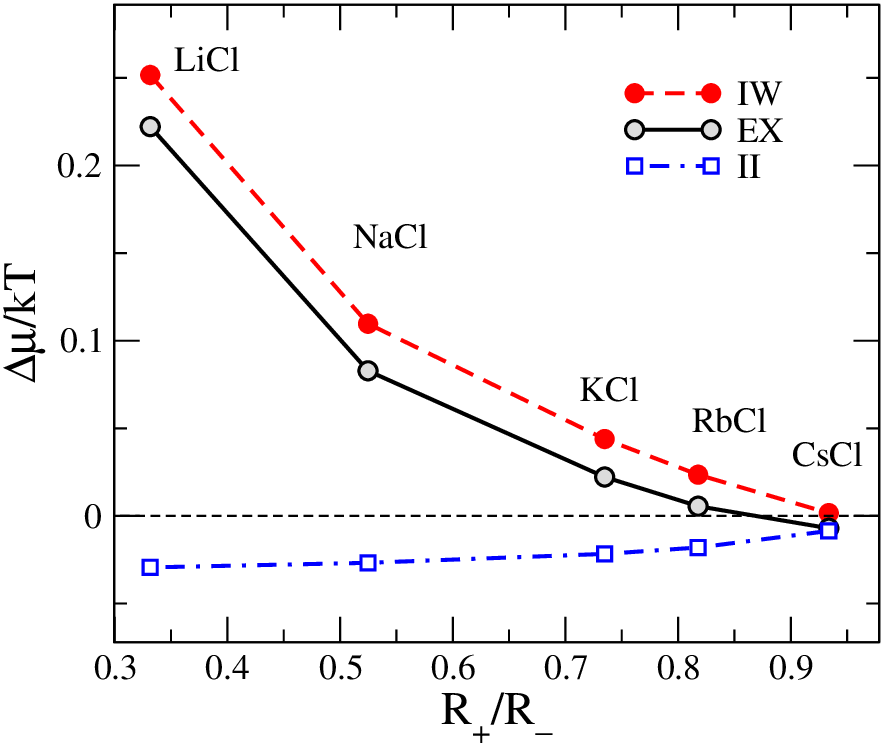}
\end{center}
\caption{The differences of the excess chemical potential components of cations and anions (see Eq.\ \ref{eq:delta_mu}) in alkali metal chlorides as functions of the ratio of the Pauling radii, $R_{+}/R_{-}$ for fixed concentration $c=1$ M. 
Solid, dot-dashed, and dashed lines refer to the $\Delta \mu^{\mathrm{EX}}(c)$, $\Delta \mu^{\mathrm{II}}(c)$, and $\Delta \mu^{\mathrm{IW}}(c)$ data, respectively. 
}
\label{fig7}
\end{figure}

The $\Delta\Delta G^{\mathrm{s}}$ difference is usually negative because the cation is either smaller than the anion (exceptions are the fluorides)  or it is double charged (see inset). 
This makes the $\Delta\mu^{\mathrm{IW}}$ difference positive (even more so in the 2:1 case).
The $R_{+}/R_{-}$ ratio is usually smaller than 1.
This makes the $\Delta\mu^{\mathrm{II}}$ difference negative because the smaller cation has stronger interaction with its ionic cloud.
It is even more pronounced in the 2:1 case, because the divalent cation has stronger interaction with its ionic cloud.
These two quantitities represent two competing effects with similar magnitudes but opposite signs.
The net effect, therefore, shows the observed behavior, which is qualitatively correct in most cases.

The explanation of the fact that we have quite good agreement in the 2:1 case is that we have an additional effect, the charge asymmetry of the ions, that is taken into account relatively correctly in our treatment.
In the 1:1 case, on the other hand, the two competing effects are quite similar, originating only from two factors, size asymmetry of the ions ($R_{+}/R_{-}$) and the difference in their solvation properties ($\Delta\Delta G^{\mathrm{s}}$).
Because our models are approximate, any error in the II and IW terms separately can add up to a considerable error in their sum.
We believe that this is the basic explanation of the relatively weak performance of our theory in the 1:1 case.

\begin{figure}[t]
\begin{center}
\includegraphics*[width=0.6\linewidth]{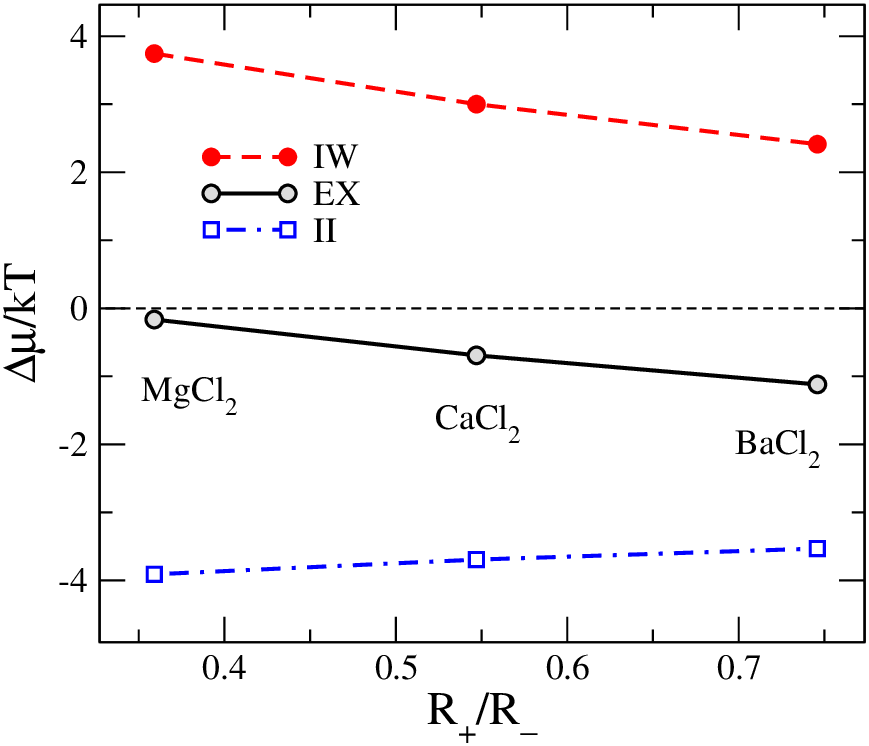}
\end{center}
\caption{The differences of the excess chemical potential components of cations and anions (see Eq.\ \ref{eq:delta_mu}) in alkali earth metal chlorides as functions of the ratio of the Pauling radii, $R_{+}/R_{-}$ for fixed concentration $c=1$ M. 
Solid, dot-dashed, and dashed lines refer to the $\Delta \mu^{\mathrm{EX}}(c)$, $\Delta \mu^{\mathrm{II}}(c)$, and $\Delta \mu^{\mathrm{IW}}(c)$ data, respectively. }
\label{fig8}
\end{figure}

To further analyze this competition, we plot the $\Delta\mu^{\mathrm{EX}}$, $\Delta\mu^{\mathrm{II}}$, and $\Delta\mu^{\mathrm{IW}}$ values for alkali metal chlorides (Fig.\ \ref{fig7}) and alkaline earth metal chlorides (Fig.\ \ref{fig8}) as functions of $R_{+}/R_{-}$ with $c=1$ M.
The II term is quite small in the 1:1 case, because only the size asymmetry of the ions causes this difference.
The dominant term is the IW term, which might give the impression that this is the term that casues the disagreement (in some cases) with experiments.
This, however, can be a false impression. 
The II term can also contain an error that can result in a deviation in the order of magnitude indicated by the figure ($<0.2 kT$).
The different stability of the hydration shells of the two ions, for example, can play an important role, but this is beyond the capabilities of our modeling level.
In the 2:1 case, on the other hand, both terms are large in order of magnitude, varying between $-4kT$ and $4kT$ (Fig.\ \ref{fig8}).

Figure \ref{fig7} also shows why $\Delta\mu^{\mathrm{EX}}$ is small for KCl, RbCl, and CsCl.
For these electrolytes, both the II and IW terms are small.
Figure \ref{fig6} shows that these electrolytes are close to the crosspoint of the two dashed line (the $R_{+}/R_{-}=1$ and $\Delta\Delta G^{\mathrm{s}}=0$ point).
Because the cation and the anion in these electrolytes are very close in nature from both points of view (II interactions and solvation), it is reasonable to assume that the activity coefficients for them are very close.
This reasoning supports Hurlen's findings (Fig.\ \ref{fig1}).

However, the data of the VWV group contradict this result.
Inspecting the data of VWV (Fig.\ \ref{fig1}), one sees that $\gamma_{i}$ of Na$^{+}$ is larger than that of anions for sodium halides, while the opposite relation is seen for potassium halides (here the $\gamma_{i}$ of K$^{+}$ is smaller).
This seems to be quite a large change considering that the only difference between the two experiments is that Na$^{+}$ is replaced with K$^{+}$.

We do not want to speculate about possible reasons of errors in any of the experiments, but in the case of KCl and similar systems, we would like to emphasize that we are talking about small effects in these cases.
Any uncertainty in any of the terms of Eq.\ \ref{eq:emf} (especially in the treatment of the junction potential) can lead to errors in $\Delta\mu^{\mathrm{EX}}$.
This is exactly the reason why we tend to trust the experimental data for electrolytes of  cations and anions that are very different.
These systems are the 2:1 electrolytes and the lithium halides, for which we have relatively good agreement with experiments. 


\section{Conclusion}
\label{sec:summary}

Our main purpose here was to understand physical mechanisms behind the behavior of the individual activity coefficients.
It was not a goal of this study to produce accurate quantitative agreement with the help of adjustable model parameters.
Our model is admittedly crude neglecting many details, but the major energy terms corresponding to basic interactions are included.
We think that our simple arguments are justified by the fact that we can produce qualitative agreement with experiments without any adjustable parameter, especially in the 2:1 case, where an anomalous behavior is reproduced.
Such an anomalous behavior is usually the result of competing energy terms.

Our definition of the II and IW terms is approximate, but not arbitrary; the division is made on the basis of the different nature of the particles in question, solute and solvent.
The concentration dependence of the dielectric constant is a hard experimental fact that should be built into theories because it has obvious large effects.
The way we do it in our II+IW model is one way, but more accurate models can and should be used.
The concentration-dependent dielectric constant is a crucial quantity coupling the calculation of the II and IW terms that otherwise can be computed independently.

Naturally, there are errors in both the II and IW terms, because the implicit solvent model used to describe them is simplistic.
These errors seem to have a larger effect in the 1:1 case, where the II and IW terms are relatively small.
In the 2:1 case, on the other hand, charge asymmetry of the ions causes large deviations between the terms for the two ions.
Lastly, we believe that the ``solvated ionic radius'' is a concept that should not be used in the description of these phenomena.
Contact positions of the ``bare'' ions are so important that they should not be excluded from the statistical sample.
The solvated ion, where the ion is moving together with strongly attached water molecules, on the other hand, is a useful idea in other cases such as transport phenomena.

\textbf{Support of experiments????}

\section*{Acknowledgment}
\label{sec:acknow}

We acknowledge the support of the Hungarian National Research Fund (OTKA NN113527) in the framework of ERA Chemistry and the J\'anos Bolyai
Research Fellowship.
Present publication was realized with the support of the projects T\'AMOP-4.2.2/A-11/1/KONV-2012-0071 and T\'AMOP-4.1.1/C-12/1/KONV-2012-0017.
The authors grateful to Dirk Gillespie for useful discussion and wise advices.


\end{document}